\documentclass{article}

\usepackage{epsfig} 
\usepackage{amsmath}
\usepackage{amsfonts}
\usepackage{graphicx}

\date{\today}

\newcommand{\insertplot}[5]{\begin{figure}
 \hfill\hbox to 0.05in{\vbox to #5in{\vfill
 \inputplot{#1}{#4}{#5}}\hfill}
 \hfill\vspace{-.1in}
 \caption{#2}\label{#3}
 \end{figure}}
 \newcommand{\inputplot}[3]{
 \special{ps: plotfile #1}
}
\newcounter{fig}

\newcommand{\beq}{\begin{equation}}
\newcommand{\eeq}{\end{equation}}
\newcommand{\beqs}{\begin{eqnarray}}
\newcommand{\eeqs}{\end{eqnarray}}

\numberwithin{equation}{section}
\newcommand{\be}{\begin{equation}}
\newcommand{\ee}{\end{equation}}
\newcommand{\bea}{\begin{eqnarray}}
\newcommand{\eea}{\end{eqnarray}}

\begin{document}

\title{$d\geq 5$ magnetized static, balanced black holes 
\\
with $S^2\times S^{d-4}$ event horizon topology} 
  
 \author{
 {\large Burkhard Kleihaus}, {\large Jutta Kunz} 
and {\large Eugen Radu}  
\\ 
\\
{\small Institut f\"ur Physik, Universit\"at Oldenburg, Postfach 2503
D-26111 Oldenburg, Germany}
}
\maketitle 
 
\begin{abstract}
We construct static, nonextremal black hole solutions of the
Einstein-Maxwell equations in $d=6,7$
spacetime dimensions, 
with an event horizon of $S^2\times S^{d-4}$ topology.
These configurations are asymptotically flat, the  $U(1)$
field being purely magnetic, with a spherical
distribution of monopole charges but no net charge measured at infinity.
They can be viewed as generalizations of the 
$d=5$ static dipole black ring, sharing its basic properties,
in particular the presence of a conical singularity.
The magnetized version of these solutions is constructed
by applying a Harrison transformation,
which puts them into an external magnetic field.
For  $d=5,6,7$, balanced configurations  approaching asymptotically
a Melvin universe background are found
for a critical value of the background magnetic field.
\end{abstract}

\section{ Introduction}
A remarkable property of  black rings is the existence of regular configurations with gauge dipoles
that are independent of all conserved charges.
This strongly contrasts with the picture valid in $d=4$ black hole physics, and
 implies a violation of the 'no hair' conjecture 
and of the black hole uniqueness.
These aspects are clearly illustrated by the 
$d=5$ black ring found by Emparan in \cite{Emparan:2004wy},
which was the first example of a black object that is asymptotically flat,
possesses a regular horizon and is the source of a dipolar gauge field.
This exact solution of the Einstein-Maxwell-dilaton equations
has an event horizon of $S^2\times S^1$ topology. 
The $U(1)$ field is purely magnetic, being produced by a circular
distribution of magnetic monopoles\footnote{The electric dual of these solutions 
can be considered as well, the ring being sourced in this case by an
electric two-form potential.}.
Then the ring creates a dipole field only,
with no net charge measured at 
infinity\footnote{Note that, as discussed in \cite{Copsey:2005se}, \cite{Astefanesei:2005ad},
the  dipole moment enters the first law of thermodynamics.}.
Similar to  the vacuum case \cite{Emparan:2001wn}, the generic dipole rings  
(in particular the static ones) are plagued
by conical singularities.
The balance is achieved for a critical (nonzero) value of the angular momentum only. 

It is clear that the dipole ring solution in \cite{Emparan:2004wy}
should have generalizations in more than five dimensions. 
However, the analytic construction of these solutions seems to be intractable within a nonperturbative approach.
Some progress in this direction has been achieved by using the blackfold approach.
There the central assumption is that some black objects, in certain ultra-spinning regimes, can
be approximated by very thin black strings or branes curved into a given shape, see 
\cite{Emparan:2007wm}, \cite{Emparan:2009cs}, \cite{Emparan:2009at}.
 Ref. \cite{Caldarelli:2010xz} has found in this way generalizations of the dipole black ring
for  several topologies of the horizon, in particular for the ring case, $S^1\times S^{d-3}$.
However,
the blackfold approach has some limitations; for example, black holes with no black membrane behavior
cannot be described within this framework.

A different approach for the 
construction of $d \geq 5$  black objects with a nonspherical topology of the
horizon has been proposed in 
 Ref. \cite{Kleihaus:2009wh}, \cite{Kleihaus:2010pr}.
The solutions are found in this case nonperturbatively, 
by solving numerically the Einstein
equations with suitable boundary conditions.
A number of new solutions 
have been constructed in this manner, in particular recently 
 Ref. \cite{Kleihaus:2012xh} has  given numerical evidence for the existence
of balanced spinning vacuum black rings  in $d\geq 6$ dimensions
beyond the blackfold limit,  and analyzed their basic properties. 
  
In this work 
we propose to construct new static nonextremal black objects with 
 a $S^2\times S^{d-4}$
topology of the event horizon in $d=6$ and $7$ dimensions,
by extending the results in \cite{Kleihaus:2009wh}
to the case of Einstein-Maxwell theory.  
These solutions can be viewed as higher dimensional generalizations of the 
$d=5$ static dipole ring in \cite{Emparan:2004wy},
the magnetic  field being analogous to a dipole, with no 
net charge measured at infinity.
However, in the absence of rotation, these configurations have a conical singularity which
provides the force balance that allows for their existence for any $d\geq 5$.
 
However, as discussed in \cite{Ortaggio:2004kr},
 the conical 
  singularity of the $d=5$ static dipole ring can be removed 
by "immersing" it in a background gauge field.
In this work we show that this holds  for $d>5$ solutions as well.
By applying
a magnetic Harrison transformation,
the conical singularities disappear for a critical value of the background magnetic field.
The resulting configurations describe $d> 5$ balanced black holes with a  horizon of $S^2\times S^{d-4}$
topology, in  a Melvin universe background.

\section{The model and general relations}
 
\subsection{The ansatz and equations}

We consider the Einstein-Maxwell theory in $d$-spacetime dimensions, defined by the following action
\begin{eqnarray}
\label{action}
S=\frac{1}{16 \pi}
\int d^dx \sqrt{-g}
\left(
{\mathcal R}-\frac{1}{4} F^2
\right) ,
\end{eqnarray}
the corresponding equations of motion being
\begin{eqnarray}
\label{eqs}
E_i^j=R_{i}^j-\frac{1}{2}\delta_i^j R 
-\frac{1}{2}(F_{ik}F^{jk}-\frac{1}{4}\delta_i^j F^2)=0,
~~
\frac{1}{\sqrt{-g}}\partial_i (\sqrt{-g}F^{ij})=0.
\end{eqnarray}
The solutions in this work are 
 static and axisymmetric configurations,
 with a symmetry group $R_t\times U(1)\times SO(d-3)$ (where $R_t$ denotes the time translation). 
Following the Appendix C of \cite{Kleihaus:2010pr}, we take the following metric ansatz:
\begin{eqnarray}
\label{metric}
 ds^2=f_1(r,\theta)(dr^2+r^2 d\theta^2)+f_2(r,\theta) d\psi^2+f_3(r,\theta) d\Omega_{d-4}^2-f_0(r,\theta) dt^2,
\end{eqnarray}
where $d\Omega^2_{d-4}$ is the unit metric on $S^{d-4}$, the range of $\theta$ is $0\leq\theta\leq \pi/2$ and  
$\psi$ is an angular coordinate, with $0\leq \psi \leq 2 \pi$. Also,  $r$ and $t$ correspond  to the radial and time
coordinates, respectively. We shall see that for the solutions in this work, 
the range of $r$ is $0<r_H\leq r<\infty$;
thus the $(r,\theta)$ coordinates have a rectangular boundary well suited for numerics.

For any value of $d$, the U(1) potential has a single component,
\begin{eqnarray}
\label{ansatz-U1}
A=A_{\psi}(r,\theta) d\psi.
\end{eqnarray}
It is of interest to mention that the model admits a dual formulation, 
with an 'electric' version of
 (\ref{action}), with
 \begin{eqnarray}
 \label{action1}
S=\frac{1}{16 \pi}
\int d^dx \sqrt{-g}
\left(
R -\frac{1}{2(d-2)!} \tilde F_{(d-2)}^2
\right),
 \end{eqnarray}
 where $\tilde F=\star F=dB$ is a $(d-2)$-form field strength 
(then the only nonvanishing components of the  $(d-3)$-form potential $B$ are $B_{\Omega t}$).
However, in this work we shall restrict to the magnetic description within the
 Einstein-Maxwell theory.

An appropriate combination 
of the Einstein equations,
 $E_t^t=0,~E_r^r+E_\theta^\theta=0$, $E_{\psi}^{\psi}=0$,
 and  $E_{\Omega}^{\Omega}=0$,
yields the following set of equations for the functions $f_1,~f_2,~f_3$ and $f_0$:
\begin{eqnarray}
\nonumber
&&\nabla^2 f_1-\frac{1}{f_1}(\nabla f_1)^2-(d-4)(d-5)\frac{f_1}{4f_3^2}(\nabla f_3)^2-\frac{f_1}{2f_0f_2}(\nabla f_0)\cdot( \nabla f_2)
\\
\nonumber
&&{~~~~~~~~}- \frac{(d-4)f_1}{2f_0f_3}(\nabla f_0)\cdot( \nabla f_3)- \frac{(d-4)f_1}{2f_2f_3}(\nabla f_2)\cdot( \nabla f_3)+\frac{(d-4)(d-5)f_1^2}{ f_3}
+\frac{(d-4)f_1}{2(d-2)f_2}(\nabla A_\psi)^2=0,
\\
\label{eqs1} 
&&\nabla^2 f_2-\frac{1}{2f_2}(\nabla f_2)^2+\frac{1}{2f_0}(\nabla f_0)\cdot( \nabla f_2)+\frac{(d-4)}{2f_3}(\nabla f_2)\cdot( \nabla f_3)
+\frac{d-3}{d-2}(\nabla A_\psi)^2=0,
\\
\nonumber
&&\nabla^2 f_3+\frac{(d-6)}{2f_3}(\nabla f_3)^2+\frac{1}{2f_0}(\nabla f_0)\cdot( \nabla f_3)
+\frac{1}{2f_2}(\nabla f_2)\cdot( \nabla f_3)-2(d-5)f_1
-\frac{f_3}{(d-2)f_2}(\nabla A_\psi)^2=0,
\\
\nonumber
&&\nabla^2 f_0-\frac{1}{2f_0}(\nabla f_0)^2+\frac{1}{2f_2}(\nabla f_0)\cdot( \nabla f_2)+\frac{(d-4)}{2f_3}(\nabla f_0)\cdot( \nabla f_3)
-\frac{f_0}{(d-2)f_2}(\nabla A_\psi)^2=0.
\end{eqnarray}
From  the Maxwell equations, it follows that the magnetic potential $A_\psi$
is a solution of the equation
\begin{eqnarray}
\label{eqs2} 
\nabla^2 A_\psi+\frac{1}{2f_0} (\nabla f_0)\cdot( \nabla A_\psi)
+\frac{1}{2f_2} (\nabla f_2)\cdot( \nabla A_\psi)
+\frac{(d-4)}{2f_3} (\nabla f_3)\cdot( \nabla A_\psi)=0.
\end{eqnarray}
In the above relations, we have defined 
$
(\nabla U) \cdot (\nabla V)=\partial_r U \partial_r V+ \frac{1}{r^2}\partial_\theta U \partial_\theta V,
$
and
$
\nabla^2 U=\partial_r^2U+\frac{1}{r^2}\partial_\theta^2 U+\frac{1}{r}\partial_r U.
$

The remaining Einstein equations $E_\theta^r=0,~E_r^r-E_\theta^\theta=0$
yield two constraints. Following \cite{Wiseman:2002zc}, we note that
setting $E^t_t=E^{\varphi}_{\varphi} =E^r_r+E^\theta_\theta=0$
in the identities
$\nabla_\mu E^{\mu r}=0$ and $\nabla_\mu E^{\mu \theta}=0$, 
we obtain the Cauchy-Riemann relations
$
\partial_\theta\left(\sqrt{-g} E^r_\theta \right) +
  \partial_{\bar r}\left(  \sqrt{-g} \frac{1}{2}(E^r_r-E^\theta_\theta) \right)
= 0 ,~~
 \partial_{\bar r}\left(\sqrt{-g} E^r_\theta \right)
-\partial_\theta \left(  \sqrt{-g} \frac{1}{2}(E^r_r-E^\theta_\theta) \right)
~= 0 , 
$
(with $r^2\partial/\partial r=\partial/\partial \bar r$).
Thus the weighted constraints satisfy Laplace equations,
and the constraints are fulfilled,
when one of them is satisfied on the boundary 
and the other at a single point
\cite{Wiseman:2002zc}. 

We close this part by remarking that the solutions in this work can also be studied by using
Weyl-like coordinates, with $ds^2=\bar f_1(\rho,z)(d\rho^2+ dz^2)+f_2(\rho,z) d\psi^2+f_3(\rho,z) d\Omega_{d-4}^2-f_0(\rho,z) dt^2,$
and $A=A_{\psi}(\rho,z) d\psi$.
The general transformation between $(\rho,z)$ and $(r,\theta)$ coordinates is given in Ref. \cite{Kleihaus:2010pr}.
Indeed, the vacuum limit of the solutions in this work ($A_\psi \equiv 0$) 
was studied in Ref. \cite{Kleihaus:2009wh} by employing the $(\rho,z)$-coordinates.
The metric Ansatz (\ref{metric}) in terms of $(r,\theta)$ allows, however, for a better numerical accuracy.

\subsection{Black holes with $S^2\times S^{d-4}$ topology of the event 
horizon}
 
\subsubsection{Boundary conditions}
The equations (\ref{eqs1}) are solved subject to a set of boundary conditions
which results from the requirement that the solutions describe asymptotically flat
black objects with a regular horizon of $S^2\times S^{d-4}$ 
topology\footnote{In obtaining these conditions we are also guided by the $d=5$ 
exact solution discussed below.}.
We assume that as $r\to \infty$,  the Minkowski spacetime background 
(with $ds^2=dr^2+r^2(d\theta^2+\cos^2\theta d\psi^2+\sin^2 \theta d\Omega_{d-4}^2)-dt^2$)
is recovered, while the gauge potential vanishes. This implies 
\begin{eqnarray}
\label{bc-inf}
f_0|_{r=\infty}=1,~ f_1|_{r=\infty}=1,~ \lim_{r\to \infty} \frac{f_2}{r^2}=\cos^2\theta,
\lim_{r\to \infty} \frac{f_3}{r^2}=\sin^2\theta,~A_\psi|_{r=\infty}=0.
\end{eqnarray}
Also, we impose the existence of a nonextremal event horizon, which is located
 at a constant value of the radial
coordinate, $r=r_H>0$.
There we require 
\begin{eqnarray}
\label{bc-rh}
f_0|_{r=r_H}=0,~\partial_r f_1|_{r=r_H}=\partial_r f_2|_{r=r_H}=\partial_r f_3|_{r=r_H}=0,~~
\partial_r A_\psi|_{r=r_H}=0.
\end{eqnarray}
The boundary conditions at $\theta=\pi/2$ are
\begin{eqnarray}
\label{bc-pi2}
\partial_\theta f_0|_{\theta=\pi/2}
=\partial_\theta f_1|_{\theta=\pi/2}
= f_2|_{\theta=\pi/2}
=\partial_\theta f_3|_{\theta=\pi/2}= 0,~~A_\psi|_{\theta=\pi/2}=0.
\end{eqnarray}
The absence of conical singularities requires  also
$r^2 f_1=f_2$ on that boundary.

The boundary conditions for $\theta=0$ are more complicated, since they encode the non-trivial topology of the horizon.
The idea here is that for some interval $r_H\leq r<R$, 
we have for the metric the same conditions as for $\theta=\pi/2$, the asymptotic behaviour
$f_2\sim \cos^2\theta$, $f_3\sim \sin^2 \theta$ being recovered for $r> R$ (with $R>r_H$ 
an input parameter).
Therefore, 
for $r_H<r<R$, we impose 
\begin{eqnarray}
\label{bc-01}
\partial_\theta f_0|_{\theta=0}
=\partial_\theta f_1|_{\theta=0}
=f_2|_{\theta=0}
=\partial_\theta  f_3|_{\theta=0}= 0,
~~ A_\psi|_{\theta=0}=\Psi.
\end{eqnarray}
For $r>R$ we require instead
\begin{eqnarray}
\label{bc-02}
\partial_\theta f_0|_{\theta=0}
=\partial_\theta f_1|_{\theta=0}
=\partial_\theta f_2|_{\theta=0}
=f_3|_{\theta=0}=0,~~
\partial_\theta A_\psi|_{\theta=0}
=0.
\end{eqnarray}
Although the constants $R,r_H$ which enter the above relations have no invariant meaning,
they provide a rough measure for the radii of the $S^{d-4}$  and $S^{2}$ spheres, respectively,  on the horizon.
Also, we shall see that the parameter $\Psi$ fixes the local charge of the solutions.

\subsubsection{Global quantities}
The  metric of a spatial cross-section of the horizon is
\begin{eqnarray}
d\sigma^2=f_1(r_H,\theta)r^2_H d\theta^2+f_2(r_H,\theta)d\psi^2+f_3(r_H,\theta)d\Omega_{d-4}^2.
\end{eqnarray}
Since, from the above boundary conditions, the orbits of $\psi$ shrink  to zero at $\theta=0$ and $\theta=\pi/2$ while 
the area of $S^{d-4}$ does not vanish anywhere, the topology of the horizon 
is $S^2\times S^{d-4}$ (in fact, for all  nonextremal  solutions in this work, $f_2(r_H,\theta)\sim \sin^2 2\theta$
while $f_1(r_H,\theta)$ and $f_3(r_H,\theta)$ are strictly positive and finite functions).
The
event horizon area is given by
\begin{eqnarray}
\label{AH}
A_H=2\pi r_H V_{d-4} \int_{0}^{\pi/2} d\theta \sqrt{f_1 f_2f_3^{d-4}}\bigg|_{r=r_H},
\end{eqnarray}
where $V_{d-4}$ is the area of the unit sphere $S^{d-4}$.

%
The Hawking temperature as computed from the surface gravity  or by requiring regularity 
on the Euclidean section, is
\begin{eqnarray}
\label{TH}
T_H=
\frac{1}{2 \pi }\lim_{r\to r_H}\sqrt{\frac{f_0}{(r-r_H)^2f_1}}=\frac{1}{\beta}~,
\end{eqnarray}
where the constraint equation $E_r^\theta=0$ guarantees that the Hawking temperature is constant on the event horizon.

At infinity, the Minkowski background is approached.
The total mass of the solutions is given by \cite{Myers:1986un}
(where the integral is taken over the $(d-2)-$sphere at spatial infinity and $k=\partial/\partial t$)
\begin{eqnarray}
\label{new-mass}
M=-\frac{(d-2)}{(d-3)}\frac{1}{16\pi}\oint_{\infty}dS_{ij} \nabla^{i} k^{j},
\end{eqnarray}
and  
can be read from the asymptotic
expression for $f_0$,
\begin{eqnarray}
\label{gtt}
-g_{tt}=f_0\sim 1-\frac{16 \pi G M}{(d-2)V_{d-2} }\frac{1}{r^{d-3}}+\dots~.
\end{eqnarray}
Using Gauss' theorem, 
the Einstein equations
and the  boundary conditions (\ref{bc-inf})-(\ref{bc-02}),
one finds from 
(\ref{new-mass})
the following Smarr-type relation
\begin{eqnarray}
\label{smarr}
(d-3)M=(d-2)\frac{1}{4}T_H A_H+ \Phi {\cal Q}.
\end{eqnarray}
Here ${\cal Q}$ is the 'local' magnetic charge which enters the thermodynamics\footnote{The asymptotic behaviour of the  magnetic potential  
is $A_{\psi}\to  {Q_{(\infty)}\cos^2 \theta}/{r^{d-3}}$.
However, $Q_{(\infty)}$ does $not$ enter any global law 
(this holds also for the $d=5$ balanced solution in \cite{Emparan:2004wy}).} as defined by 
evaluating the magnetic flux
over the $S^2$ sphere around the horizon,
\begin{eqnarray}
\label{Q}
 {\cal Q}= \frac{1}{4 \pi}\int_{S^2}F_{\theta \psi}d \theta d\psi=-\frac{\Psi}{2},
\end{eqnarray}
and $\Phi$ is the thermodynamical conjugate variable to ${\cal Q}$,
\begin{eqnarray}
\label{Phi}
\Phi= \frac{1}{8\pi}\int_0^{2\pi} d\psi \int d\Omega_{d-4 }\int_{r_H}^R dr \sqrt{-g}F^{\theta \psi}\bigg|_{\theta=0}=
\frac{1}{4}V_{d-4}\int_{r_H}^R \frac{dr }{r}\sqrt{\frac{f_0f_3^{d-4}}{f_2}}\partial_\theta A_{\psi}\bigg|_{\theta=0},
\end{eqnarray}
such that
 $
  \frac{1}{16 \pi}\int F^2 \sqrt{-g} d^{d-1}x= 2 \Phi {\cal Q}.
$
Therefore, following \cite{Emparan:2001rp}, we interpret the solutions as describing 
a spherical $S^{d-4}$ distribution of monopole charges, though with a zero net charge (see also \cite{Dowker:1995sg}).
  
As expected, in the absence of rotation, all these black objects with $S^2\times S^{d-4}$
horizon topology are plagued by conical singularities.
As one can see from the boundary conditions,
in this work we have chosen\footnote{It is also possible to work 
with the conical singularity stretching
towards the boundary. However, in that case the spacetime
will not be asymptotically flat.} to locate the conical singularity 
at  $\theta=0$, $r_H<r<R$, where we find a conical excess
\begin{eqnarray}
\label{delta}
\delta= 2\pi (1-\lim_{\theta\to 0}\frac{f_2}{\theta^2 r^2 f_1 })<0~.
\end{eqnarray}
This can be interpreted as the higher dimensional analogue of a `strut'
($e.g.$ a membrane for $d=5$),
preventing the collapse of the configurations.
Although the presence of a conical singularity is an undesirable feature,
it has been argued in \cite{Herdeiro:2009vd}, \cite{Herdeiro:2010aq}, that such asymptotically flat black objects still
admit a thermodynamical description. Moreover, when working with the appropriate set
of thermodynamical variables, the Bekenstein-Hawking law still holds, 
while the parameter $\delta$
enters the first law of thermodynamics.
Without going into details, we mention that the conjugate extensive variable to  $ {\delta}$ is
\begin{eqnarray}
\label{A}
{\cal A}\equiv \frac{Area} {\beta},
\end{eqnarray}
 where $Area$ is the space-time area of the conical singularity's world-volume.
 For the line-element (\ref{metric}), 
 the line element of the two dimensional surface spanned by the conical singularity is  
\begin{eqnarray}
\label{sigma}
d\sigma^2=-f_0dt^2+f_1dr^2+f_3d\Omega_{d-4}^2,
\end{eqnarray}
which implies
\begin{eqnarray}
\label{Area}
{\cal A}=  V_{d-4} \int_{r_H}^R dr \sqrt{f_0f_1 f_3^{d-4}}\bigg|_{\theta=0}. 
\end{eqnarray} 
 
\section{The solutions}
\subsection{The $d=5$ static dipole black ring}

The static dipole black ring  is usually written in ring or in Weyl coordinates,
where it takes a relatively simple  form. In what follows we shall write it 
within the ansatz (\ref{metric}), (\ref{ansatz-U1}), which results in rather
complicated  expressions. However, this helps us to make contact with the numerical solutions
found for $d>5$.

In  the $(r,\theta)$-coordinates, the metric functions $f_i$ in the
line element (\ref{metric}) are given by
(note that for $d=5$, the  sphere $\Omega_{d-4}$ reduces to a circle):
\begin{eqnarray}
\label{1d=5}
f_1(r,\theta)=c_2(r,\theta) f_1^{(0)}(r,\theta) ,~
 f_2(r,\theta)= \frac{f_2^{(0)}(r,\theta)} {c_1^2(r,\theta)},~
 f_3(r,\theta)= c_1(r,\theta)  f_3^{(0)}(r,\theta),~
 f_0(r,\theta)=c_1(r,\theta)f_0^{(0)}(r,\theta),~{~}
\end{eqnarray}
where
\begin{eqnarray}
\label{d=52}
c_1=1-\frac{2(R^4-r_H^4)^2}{R^4 r_H^2}\frac{w}{1+w}\frac{1}{P_{(-)}},
~~
c_2=\frac{1}{(1+w)^3}
\left(1+w\frac{R^4+r_H^4}{2R^2r_H^2}\frac{Q_{(-)}}{S_1} \right)
\left(1+w\frac{R^4+r_H^4}{2R^2r_H^2}\frac{Q_{(+)}}{S_1} \right)^2,
\end{eqnarray}
 the magnetic potential (written in a gauge such that $A_{\psi}(\theta=\pi/2)=0$) being
\begin{eqnarray}
\label{d=53}
A_\psi=\sqrt{6} \sqrt{1+\frac{R^2-r_H^2}{2R^2r_H^2}\frac{w}{1+w}}\frac{R^2-r_H^2}{R^2r_H}\frac{\sqrt{(1-w)w}}{1+w}\frac{S_{(+)}}{P_{(+)}}.
\end{eqnarray}
In the above relations we note
\begin{eqnarray}
\label{d53}
\nonumber
&&
S_{(\pm)}=r^2+\frac{r_H^4}{r^2} \pm \frac{(R^2+r_H^2)^2}{R^2}+2 r_H^2 \cos 2\theta -R_4,
~~
P_{(\pm)}=\frac{(r^2\pm r_H^2)^2(R^2+r_H^2)^2}{r^2r_H^2R^2}-2(1+\frac{R^2-r_H^2}{2R^2r_H^2}\frac{w}{1+w})S_{(\pm)},
\\
&&
Q_{(\mp)}=\frac{(r^2\pm r_H^2)^2(R^2+r_H^2)^2}{r^2(R^4+r_H^4)}-S_{(\pm)},
~~S_1=\frac{(r^2- r_H^2)^2(R^2+r_H^2)^2}{2r^2R^2r_H^2)}-S_{(-)},
\end{eqnarray}
with 
$
R_4=\sqrt{(\frac{r_H^4+R^4}{R^2}-\frac{r^4+r_H^4}{r^2}\cos 2\theta)^2+\frac{(r^4-r_H^4)^2}{r^4}\sin^22\theta}$.
Also, $f_i^{(0)}$ are the functions which enter the line element of the $d=5$
static vacuum black ring, with 
\begin{eqnarray}
\label{d=5}
f_1^{(0)}(r,\theta)=\frac{1}{F_1(r,\theta)},~~
f_2^{(0)}(r,\theta)=r^2\frac{F_2(r,\theta)}{F_3(r,\theta)} ,~~
f_3^{(0)}(r,\theta)= r^2 F_3(r,\theta),~~
f_0^{(0)}(r,\theta)= F_0(r),
\end{eqnarray}
and
\begin{eqnarray}
\nonumber
&&
F_0=(\frac{r^2-r_H^2}{r^2+r_H^2})^2,~~
F_1=\frac{R_3}{(1-\frac{r_H^2}{R^2})^2(1-\frac{r_H^2}{r^2})(1+\frac{r_H^2}{r^2})^4}
\left [
(1+\frac{r_H^4}{r^4})(1+\frac{r_H^4}{R^4})-\frac{4r_H^4}{r^2R^2}\cos 2\theta -\frac{2r_H^2}{R^2}R_3
\right ],
\\
\label{vacuumBR2}
&&
F_2=(1+\frac{r_H^2}{r^2})^4\sin^2\theta \cos^2 \theta,~~
F_3=\frac{1}{2}
\left [
R_3+\frac{R^2}{r^2}
\left(1+\frac{r_H^4}{R^4}
-\frac{r_H^2}{R^2}(\frac{r_H^2}{r^2}+\frac{r^2}{r_H^2})\cos 2\theta
\right)
\right],
 \end{eqnarray}
where
$R_3=\sqrt{(1+\frac{R^4}{r^4}-\frac{2R^2}{r^2}\cos 2\theta)(1+\frac{r_H^8}{r^4R^4}-\frac{2r_H^4}{r^2R^2}\cos 2 \theta)}$.

This solution has three parameters, $r_H,R$ (which were introduced in the previous section) and $w$,
which is fixed by the value of the magnetic potential at $\theta=0,~r_H<r<R$ via
 (note that $0\leq w<1$):
 \begin{eqnarray}
\label{Psi}
\Psi=\frac{2}{R}\sqrt{\frac{w}{1-w^2}}\sqrt{2R^2 r_H^2+(R^4+r_H^4)w)}.
\end{eqnarray}
A direct computation shows that this is indeed a solution of the Einstein-Maxwell equations. 
Also, one can see that $c_1\to 1,~c_2\to 1$ and $A_\psi\to 0$ as $w\to 0$, 
 this corresponding to the vacuum black ring limit.
 
The computation of the quantities of interest for this solution
is a straightforward application of the general formalism in Section 2.2.2.
 In the nonextremal case,  one can write the following 
 suggestive expressions:
\begin{eqnarray}
\label{cg1}
M=M^{(0)}(1+U),~~T_H=\frac{T_H^{(0)}}{(1+U)^{3/2}},~~A_H=A_H^{(0)}(1+U)^{3/2},
\end{eqnarray}
where
\begin{eqnarray}
\label{cg2}
M^{(0)}=\frac{3\pi r_H^2}{4},~~T_H^{(0)}=\frac{R^2+r_H^2}{8\pi R r_H^2},~~
A_H^{(0)}=4\pi^2 \frac{Rr_H^4}{R^2+r_H^2}
\end{eqnarray}
are the mass, temperature and area of the vacuum static black ring solution,
and
\begin{eqnarray}
\label{cg23}
{\cal Q}=\frac{2\sqrt{3}R r_H^2\sqrt{U(1+U)} }{\sqrt{R^4+r_H^4-2R^2r_H^2(1+2U)} } ,~~
\Phi=\frac{\sqrt{3}\pi}{2R}\frac{\sqrt{U(R^4+r_H^4-2R^2r_H^2(1+2U))}}{\sqrt{1+U}},
\end{eqnarray}
with $0\leq U<(R^2-r_H^2)^2/(4 R^2r_H^2)$  a free 
parameter\footnote{The relation between $w$ and $U$ is
$w=2R^2 r_H^2U/((R^2-r_H^2)^2-2R^2r_H^2 U)$.}.
The static dipole rings have a conical excess
\begin{eqnarray}
\label{cg3}
\delta=2\pi\left [ 1-\frac{R^2+r_H^2}{R^2-r_H^2}\big(1+\frac{4 R^2r_H^2U}{ R^4+r_H^4-2R^2r_H^2(1+2U)}  \big)^{3/2}\right ],
\end{eqnarray}
while the expression of the corresponding conjugate   extensive variable ${\cal A}$ cannot be written in closed form.

The basic properties of the $d=5$ non-extremal solution turn out to be generic and will be discussed in the following subsection.
Here we mention only that the extremal solutions are found by taking the limit 
$r_H\to 0$ in the relations (\ref{1d=5})-(\ref{vacuumBR2}). They have a relatively simple form
\begin{eqnarray}
\nonumber
&&
f_1=\frac{(r^2+w T_{(-)})(r^2+w T_{(+)})^2}{(1+w)^4r^4R_1},~~
f_2=\frac{2r^4 \sin^2 \theta \cos^2\theta(r^2+w T_{(+)})^2}{(T_{+}-r^2 \cos2\theta)(r^2+w T_{(+)})^2 },~~
f_3=\frac{(T_{(+)}-r^2 \cos2\theta)(r^2+w T_{(-)})}{2(r^2+wT_{(+)})},~~
\\
\label{ex1}
&&
f_0=\frac{r^2+w T_{(-)}}{r^2+w T_{(+)}},~~A_\psi=Rw\sqrt{3}\sqrt{\frac{1-w}{1+w}}\frac{r^2-T_{(-)}}{r^2+wT_{(+)}},
\end{eqnarray}
with $T_{(\pm)}=R_1\pm R^2$, $R_1=\sqrt{r^4+R^4-2r^2R^2\cos 2\theta}$
and $w={\cal Q}/\sqrt{{\cal Q}^2+3R^2}$.
The horizon of the extremal solutions has zero area, since the length of the $S^1$ direction
vanishes there, $g_{\psi\psi} \to 0$.
Their mass and and potential are given by
$M=\frac{3\pi {\cal Q} R^2}{4({\cal Q}+\sqrt{{\cal Q}^2+3R^2})}$,~
$\Phi= \frac{3\pi R^2}{2({\cal Q}+\sqrt{{\cal Q}^2+3R^2})}$,
their conical excess is $\delta=-\frac{4\pi(4{\cal Q}^3+9 R^2)}{(-{\cal Q}+\sqrt{{\cal Q}^2+3R^2})^3}$.

\subsection{$d=6,7$ numerical solutions}
\subsubsection{Remarks on the numerics}

Higher dimensional generalizations of the $d=5$ nonextremal solution (\ref{1d=5})-(\ref{vacuumBR2}) are 
found by replacing in the five dimensional line element the $S^1$ direction 
 which is  not associated with the magnetic potential,
 with the line element of a round $(d-4)$-sphere, while preserving at the same time
 the basic properties of the metric functions and of the magnetic potential.
 
Since no closed form solution is available in this case, 
 the set of five coupled non-linear
elliptic partial differential equations (\ref{eqs1}), (\ref{eqs2}) is solved numerically,
subject to the boundary conditions (\ref{bc-inf})-(\ref{bc-02}).

The numerical scheme we have used is identical with that described
at length in \cite{Kleihaus:2010pr} and thus we shall not enter into details.
We mention only that in practice we have worked with a set of 'auxiliary' functions $\mathcal{F}_i$
defined via\footnote{Note that this procedure has some similarities with the construction 
of distorted black holes \cite{Geroch:1982bv}. 
However, in our case, the field equations do not reduce  to simple Laplace equations. }
\begin{eqnarray}
\label{num1}
f_0=f_0^{(0)}e^{\mathcal{F}_0}, ~~f_1=f_1^{(0)}e^{\mathcal{F}_1},
~~
f_2=f_2^{(0)}e^{\mathcal{F}_2},~~f_3=f_3^{(0)}e^{\mathcal{F}_3},
 \end{eqnarray}
where $f_i^{(0)}$ are 'background' functions
corresponding to the $d=5$ static vacuum black ring as given by (\ref{d=5}).
These 'background' functions $f_i^{(0)}$ are used to fix the topology of the horizon
and to 'absorb' the coordinate divergencies of the functions $f_i$.
The 'auxiliary' functions $\mathcal{F}_i$ are smooth and finite everywhere 
such that they do not 
lead to the occurence of new zeros of the functions $f_i$ 
(therefore the rod structure of the solutions remains fixed by $f_i^{(0)}$ \cite{Kleihaus:2010pr}).
However, $\mathcal{F}_i$ encode the effects of changing the spacetime
dimension from $d=5$
and also of introducing the local charge ${\mathcal Q}$.

In our approach, the input parameters are the value $d$ of the spacetime dimension, 
the event horizon radius $r_H$, the radius $R$ of the
$S^{d-4}$ sphere,
and the value of the local charge ${\mathcal Q}$ ($i.e.$ the parameter $\Psi$ in the boundary conditions (\ref{bc-01})). 
 The physical parameters are
encoded in the values of
the  functions $f_i$ (and their derivatives) on the boundary of the integration domain.
For example, the mass parameter $M$ is computed from the asymptotic 
form (\ref{gtt}) of the metric function $g_{tt}=-f_0$, the Smarr relation (\ref{smarr})
being used to verify the accuracy of the solutions. 

\begin{figure}[ht]
\hbox to\linewidth{\hss%
	\resizebox{8cm}{6cm}{\includegraphics{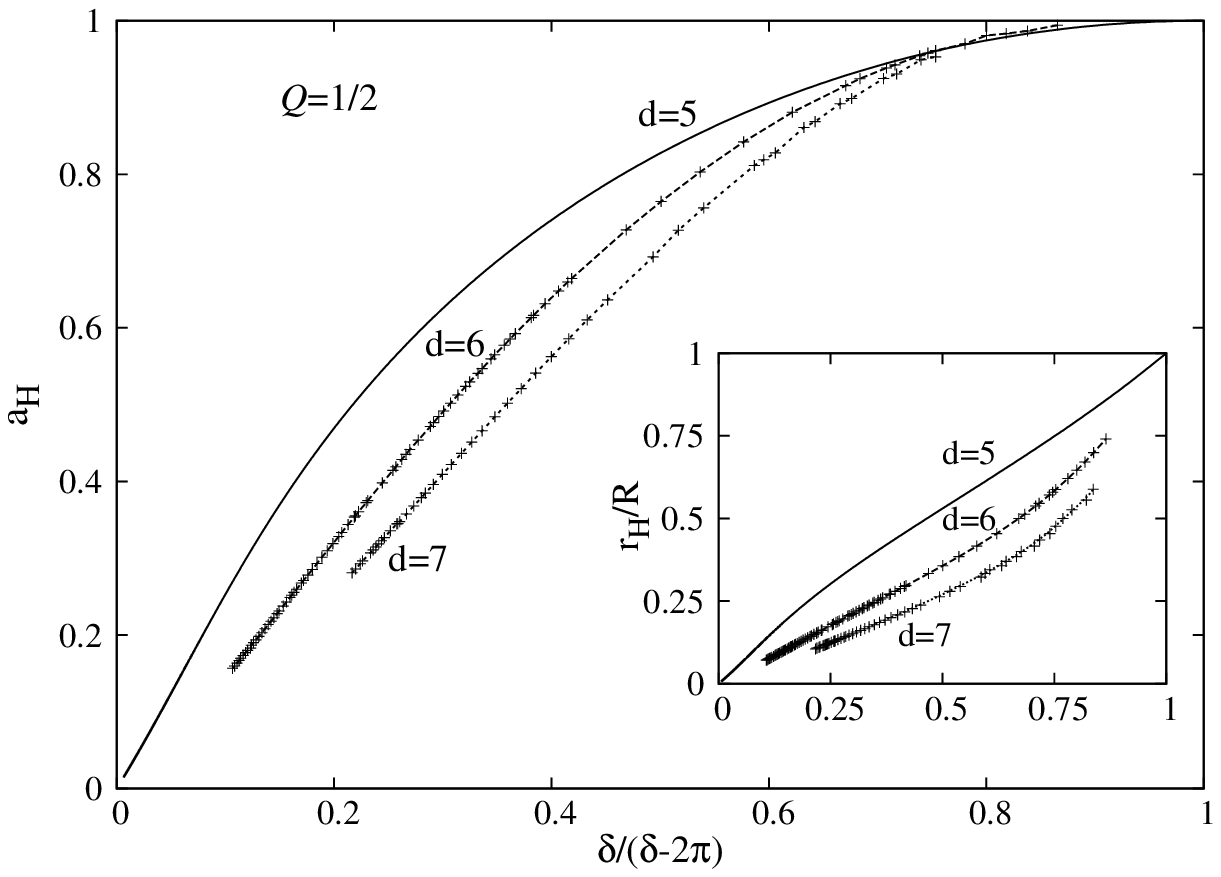}}
\hspace{5mm}%
        \resizebox{8cm}{6cm}{\includegraphics{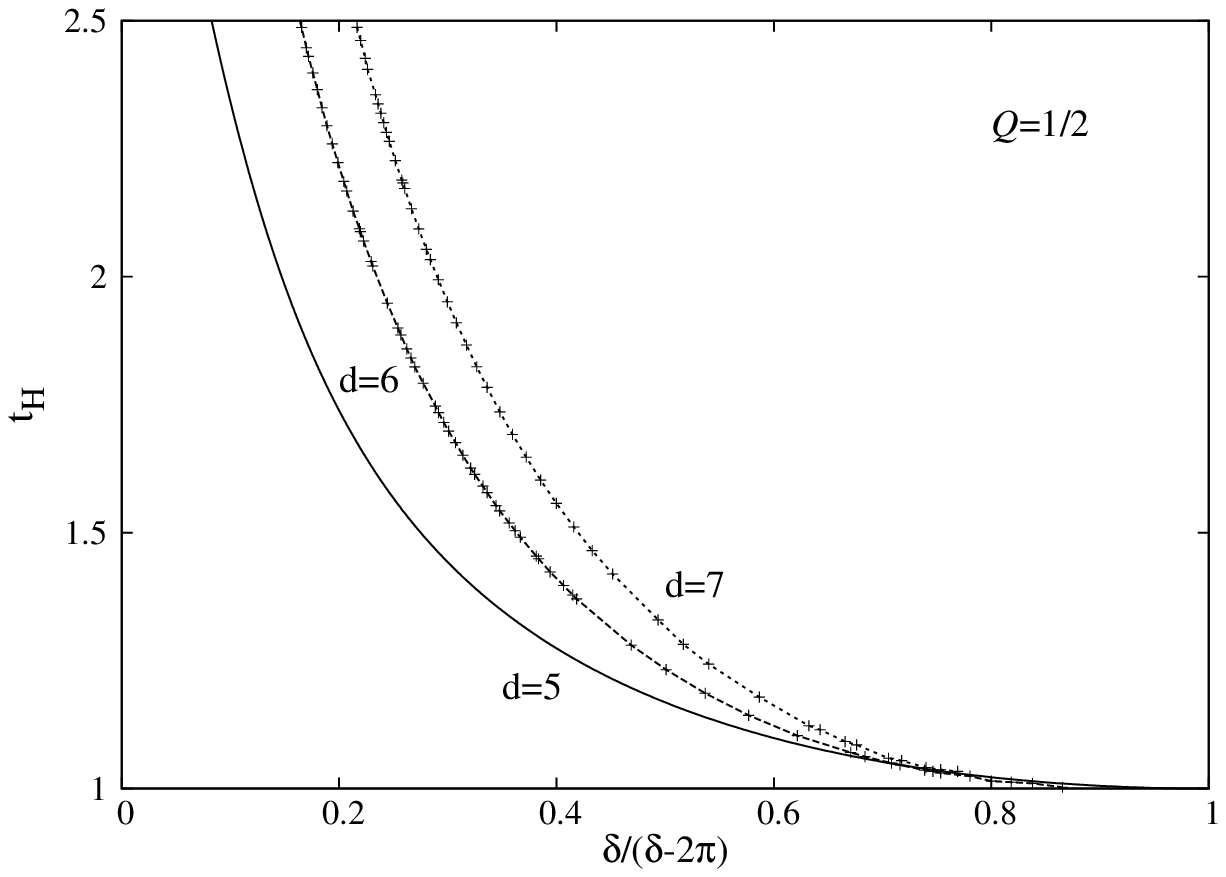}}	
\hss}
\end{figure}
\begin{figure}[ht]
\hbox to\linewidth{\hss%
	\resizebox{8cm}{6cm}{\includegraphics{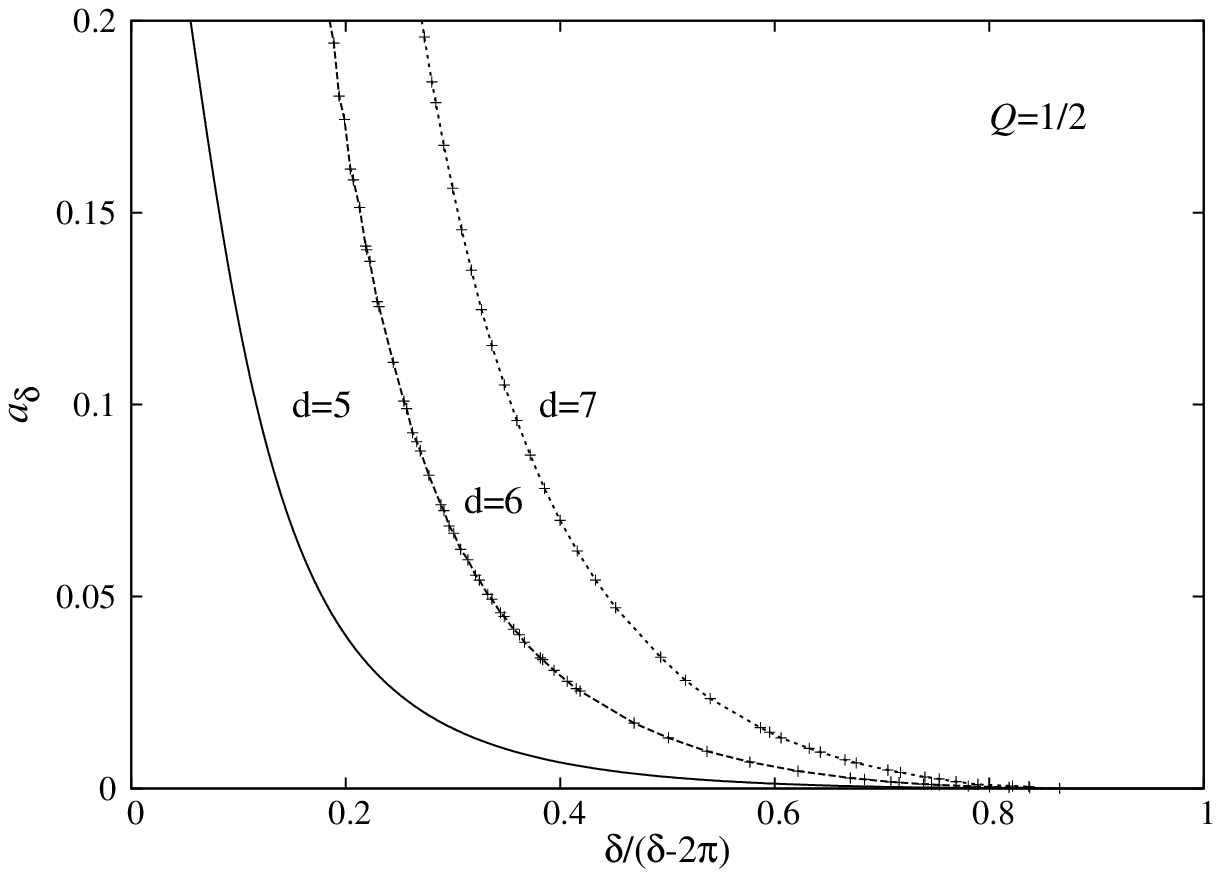}}
\hspace{5mm}%
        \resizebox{8cm}{6cm}{\includegraphics{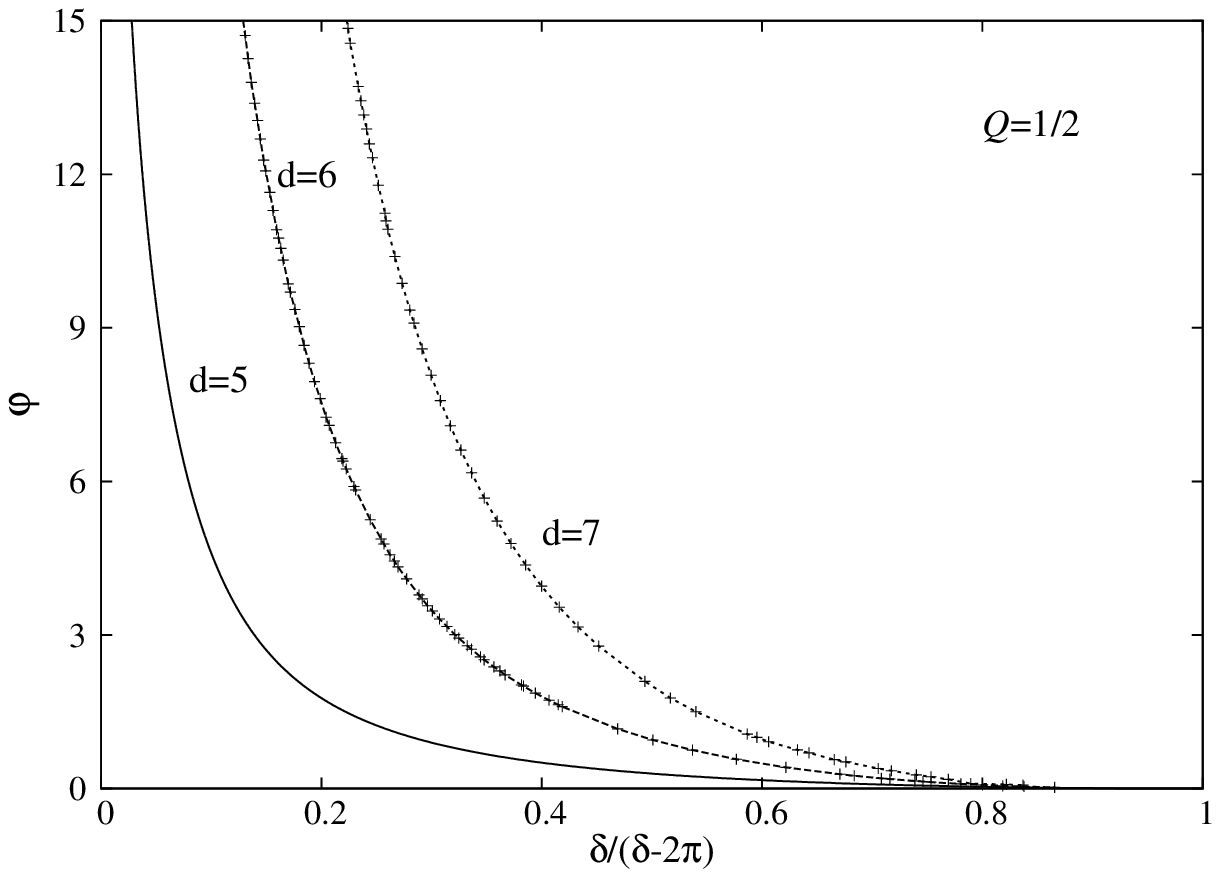}}	
\hss}\caption{\small  A number of quantities are shown as a function
of the relative angular excess $ \delta/(\delta-2\pi)$ 
for black hole solutions  with the same local charge ${\mathcal Q}$.
 }
\label{fig3}
\end{figure}

\subsubsection{Properties of the  solutions}
 
 To obtain nonextremal Einstein-Maxwell solutions with $S^2\times S^{d-4}$ horizon topology, 
one starts with the vacuum configurations in \cite{Kleihaus:2009wh} 
and turns on the parameter $\Psi$ which enters the boundary conditions for the 
magnetic potential.
The iterations converge, and, in principle, repeating the procedure it is possible
  to obtain solutions with arbitrary values of ${\mathcal Q}$.

We have started with a test of the numerical scheme, by recovering  in this way
the $d=5$ static dipole black rings.
Afterwards, new  solutions in $d=6,7$ dimensions have been studied in a systematic way. 
 Solutions with $d>7$ should also exist; 
 however, we did not try to find them and their study may require a different numerical method.  
We mention that, for all solutions, we have verified that the Kretschmann scalar stays finite
everywhere\footnote{Here we ignore the $\delta$-Dirac terms in the expression of the
Riemann tensor in the presence of a conical singularity. In fact, the presence of   a conical singularity 
has a rather neutral effect on the numerics, since the solver does not notice it directly.}.
 
 The central result in this work is that the $d=5$ static nonextremal
  dipole ring has higher dimensional generalizations with a $S^2\times S^{d-4}$ horizon topology.
 Moreover, the properties of the five dimensional solutions are generic,
 being recovered for $d>5$. 
 
 Let us start with a discussion of the solutions' features 
 for a fixed value of the magnetic charge ${\mathcal Q}$. 
 Perhaps the most important feature is that all $d\geq 5$ solutions
 have conical singularities.
  %
  %
  %
\begin{figure}[ht]
\hbox to\linewidth{\hss%
	\resizebox{8cm}{6cm}{\includegraphics{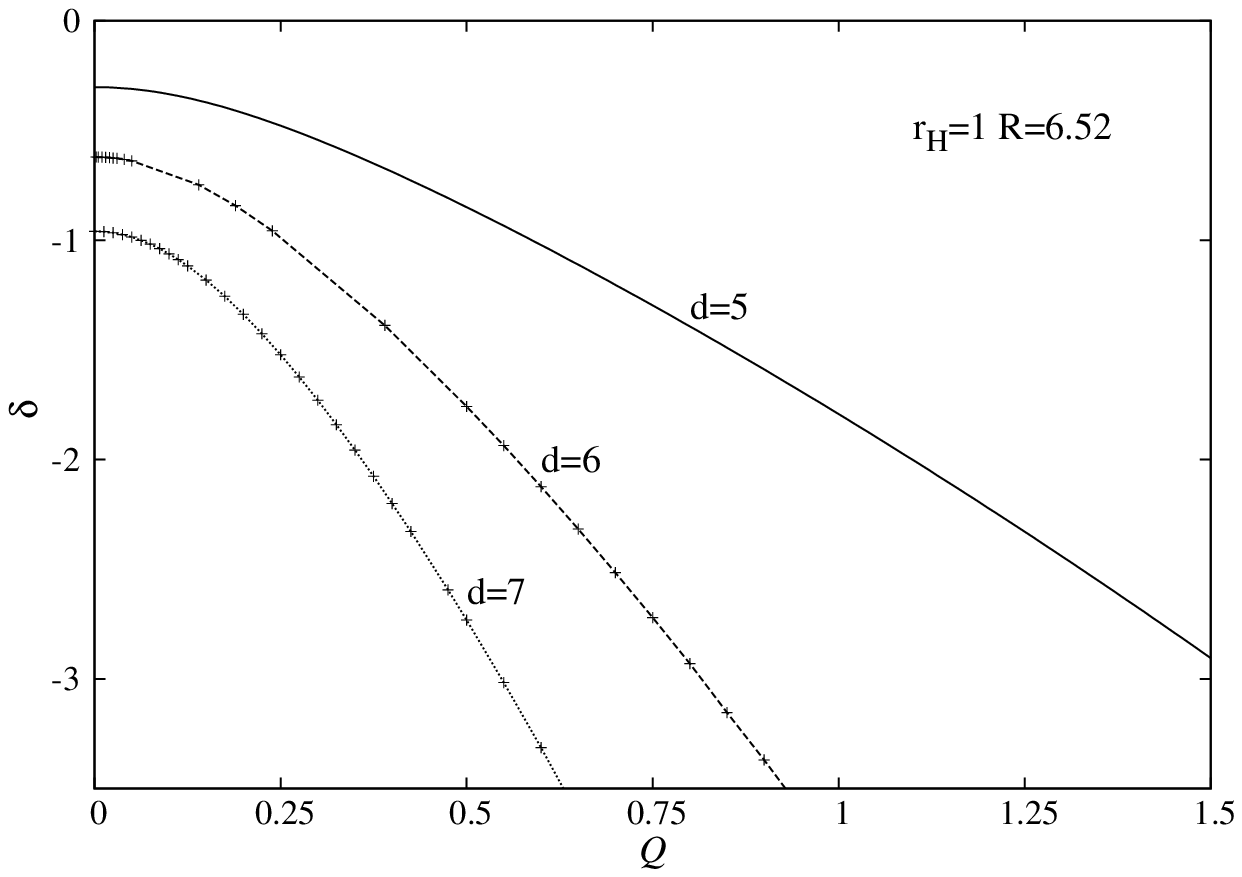}}
\hspace{5mm}%
        \resizebox{8cm}{6cm}{\includegraphics{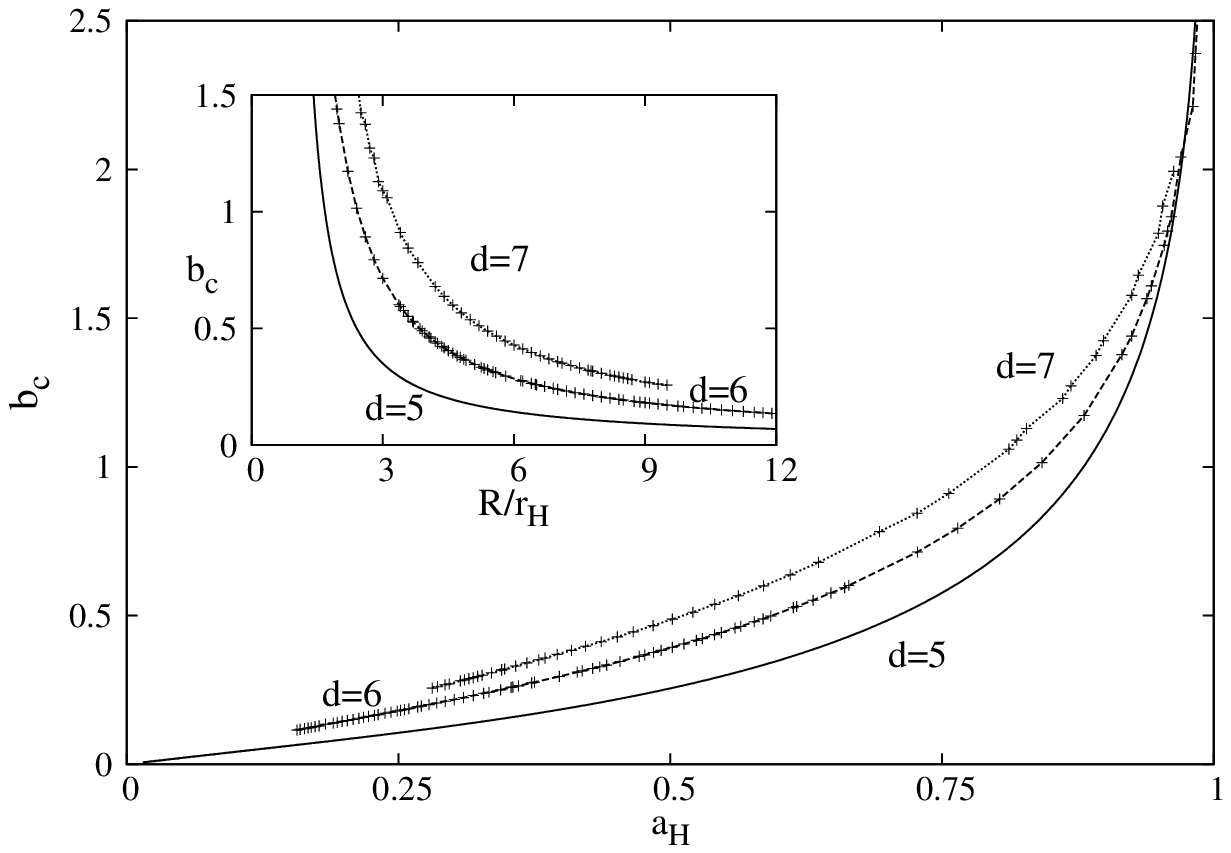}}	
\hss}\caption{\small { $Left:$} The  angular excess $\delta$ is shown as a function of the local
charge for solutions with the same values of $r_H,R$.
 { $Right$:} The  dimensionless critical magnetic  field $b_c=B {\cal Q}$ is shown as a function of 
 the dimensionless event horizon area for balanced  black holes with $S^2\times S^{d-4}$ 
 horizon topology in a Melvin universe background. The inset shows $b_c$ as a function of the 
ratio $R/r_H$.
 }
\label{fig4}
\end{figure}
%
Thus we have found it convenient to take the relative conical excess $\delta/(\delta-2\pi)$
as the control parameter and to consider the
 following dimensionless quantities\footnote{Note that in a numerical approach it is rather 
 difficult to work with dimensionless `reduced' quantities in a systematic
 way, since, except for ${\mathcal Q}$,
 it is not possible to fix  any other quantity which enters the Smarr relation and the first law.}, the scale being fixed here by $M$:
 \begin{eqnarray}
\label{reduced}
a_H=p_1 \frac{A_H}{M^{\frac{d-2}{d-3}}},~~
t_H=p_2 T_H M^{\frac{1}{d-3}},~~
a_{\delta}=\frac{1}{V_{d-4}}\frac{\cal{A}}{M}, ~~
\varphi=\frac{\Phi}{ M^{\frac{d-4}{d-3}}},
  \end{eqnarray}
  with $p_1=((\frac{ d-2 }{16\pi})^{d-2}V_{d-2})^{\frac{1}{d-3}}$,
   $p_2=\frac{1}{d-3}(\frac{2^{2(d-1)}\pi^{d-2}}{(d-2)V_{d-2}})^{\frac{1}{d-3}}$
two coefficients which 
   have been chosen such that  $a_H=1,t_H=1$  corresponds to the Schwarzschild-Tangherlini black hole.
 
In terms of the dimensionless ratio $r_H/R$, the solutions   interpolate
between two limits (although these
regions of the parameter space are difficult to approach numerically).
For $ R\to \infty$ and  $r_H,~{\cal Q}$ nonvanishing, the radius on the horizon of the 
$S^{d-4}$-sphere increases and asymptotically it becomes a $(d -4)-$plane, while $\delta \to 0$.
After a suitable rescaling\footnote{For the
$d=5$ exact solution, this rescaling is 
$r\to \sqrt{2 R r},~r_H\to \sqrt{2R r_H},~w\to w/R$.
}, one finds the  magnetically charged black brane solution
  \begin{eqnarray}
\label{brane-limit1}
&&
ds^2=H^2(r)U_1(r)\big[dr^2+r^2(  4 d \theta^2+\sin^2 2\theta d\psi^2) \big]+
\frac{1}{(H(r))^{\frac{2}{d-3}}}\big (dx_1^2+\dots+dx_{d-4}^2-U_0(r)dt^2 \big),~~{~~~}
\\
&&
\nonumber
A=-{\cal Q}(1+\cos2\theta)d\psi, 
\end{eqnarray}
 where
  \begin{eqnarray}
\label{brane-limit2}
 H(r)=\frac{1}{(r+r_H)^2}\left(r^2+r_H^2+2rr_H \sqrt{1+\frac{d-3}{8(d-2)}\frac{{\cal Q}^2}{r_H^2}}\right),~
 U_1(r)=(1+\frac{r_H}{r})^4,~U_0(r)=\big(\frac{r-r_H}{r+r_H}\big)^2.
\end{eqnarray}
This corresponds to
a magnetically charged Reissner-Nordstr\"om black hole
 uplifted to $d-$dimensions, $i.e.$ with $(d-4)$-flat directions.

The limit $r_H/R\to 1$ is somehow more subtle, since the conical excess diverges, $\delta\to -\infty$,
and the magnetic field vanishes.
As can be seen in Figure 1, 
 the Schwarzschild-Tangherlini black hole with an $S^{d-2}$ horizon topology 
 is recovered in this limit.  
This can be understood by studying the $d=5$ exact solution.
There,
as $R\to r_H$ one finds $c_1\to 1+O(R-r_H),~~c_2\to 1+O(R-r_H)$ while $A_\psi\sim O(R-r_H)^2$ ($i.e.$
a vanishing charge), with the limiting expressions
 $f_1= {f_2}/(r^2\cos^2\theta)=f_3/(r^2\sin^2\theta)=(1+r_H^2/r^2)^2,~f_0=(r^2-r_H^2)^2/(r^2+r_H^2)^2$.
  
  Some results illustrating these aspects are shown in Figure 1 (note that we have found similar results for other values
  of ${\mathcal Q}$ as well).

A different situation which can be studied numerically is to keep fixed
the radii $r_H$ and $R$ and to vary the value of the local charge ${\mathcal Q}$.
Interestingly, turning on a magnetic field increases the absolut value of the conical excess, see 
Figure 2 (left).
For fixed $r_H,R$,
the values of the magnetic potential, horizon area, the parameter ${\cal A }$ and the mass
increase with ${\mathcal Q}$, while the temperature decreases. 

It seems that similar to the $d=5$ case, the extremal solutions are found in the limit $r_H\to 0$,
for nonvanishing $R$ and ${\mathcal Q}$. 
However, we could not approach this limit and 
the numerical construction of the extremal
solutions would require a different numerical scheme, with
 another set of 'background' functions.
This holds also for the $d=5$ solutions, in which case it can be understood by noticing that
the   behaviour of the metric functions $f_1$, $f_3$ as  $r\to r_H$ 
($i.e.$ $f_1\sim 1/r^2$, $f_2\sim r^2$)  
is not compatible with the boundary conditions (\ref{bc-rh}).
We conjecture that the picture found for $d=5$ is generic and
the extremal solutions will always possess a horizon with vanishing area.

\section{Balanced black holes with
$S^2\times S^{d-4}$ event horizon topology in a Melvin universe background }

The occurrence of conical singularities is not an unusual feature in general relativity.
However, sometimes this pathology can 
be cured by placing the solutions in an 
external field (see $e.g.$ \cite{Emparan:1999au}-\cite{Yazadjiev:2008ty}).
This was the case for the $d=5$ static dipole ring \cite{Ortaggio:2004kr}
and also for extremal solutions in \cite{Emparan:2001rp},
which could  be balanced 
by "immersing" them in a background gauge field, via a magnetic Harrison transformation.
Unsurprisingly, this works also for the configurations considered in this work.

 The magnetic Harrison transformation can be summarized as follows (see $e.g.$ \cite{Gal'tsov:1998yu}).
 Let us consider a solution of the Einstein-Maxwell equations of the form
 \begin{eqnarray}
\label{h1}
ds^2=g_{yy} dy^2+d\sigma^2_{d-1},~~A=A_y dy,
\end{eqnarray}
with $\partial /\partial y$ a Killing vector.
Then the configuration
 \begin{eqnarray}
\label{h2}
ds^2=\frac{1}{\Lambda^2}g_{yy} dy^2+\Lambda^{\frac{2}{d-3}}d\sigma^2_{d-1},
~~~
A=\frac{1}{\Lambda}
\left [
A_y+B\left(g_{yy}+\frac{d-3}{2(d-2)}A_y^2 \right)
\right ] dy,
\end{eqnarray}
with
 \begin{eqnarray}
\label{Lambda}
\Lambda=(1+\frac{d-3}{2(d-2)}B A_y)^2+\frac{d-3}{2(d-2)}B^2g_{yy},
\end{eqnarray}
 solves also the Einstein-Maxwell equations (with $B$ an arbitrary parameter).
 
The Harrison   transformation (\ref{h2}) applied with respect to the Killing vector
$\partial/\partial\psi$ results
in the following line element
 \begin{eqnarray}
\label{h4}
&ds^2=\Lambda^{\frac{2}{d-3}}
\bigg(
f_1 (dr^2+r^2d\theta^2)+f_3d\Omega^2_{d-4}-f_0dt^2
\bigg) +\frac{1}{\Lambda^2}f_2 d\psi^2,~{\rm with}~
\Lambda=(1+\frac{d-3}{2(d-2)}B A_\psi)^2+\frac{d-3}{2(d-2)}B^2f_{2},~{~~~}
\end{eqnarray}
and the new magnetic potential
 \begin{eqnarray}
\label{h5}
A_\psi'=\frac{1}{\Lambda}
\left [
A_\psi+B\left(f_2+\frac{d-3}{2(d-2)}A_\psi^2 \right)
\right ].
\end{eqnarray}
One can see that the new line element preserves some of the basic properties of the 
$B=0$ seed configuration.
The horizon is still located at $r=r_H$ and has an $S^2\times S^{d-4}$ topology,
since the qualitative behaviour of the metric functions 
at $\theta=0,\pi/2$ remains unchanged (note that $\Lambda>0$ everywhere).
However, the geometry is distorted and the asymptotic behaviour is very different.
As $r\to \infty$, the solution  becomes
 \begin{eqnarray}
\nonumber
&ds^2=\Lambda^{\frac{2}{d-3}}
\left(
 dr^2+r^2(d\theta^2+\sin^2 \theta d\Omega^2_{d-4})-dt^2
\right) +\frac{r^2\cos^2 \theta }{\Lambda^2}d\psi^2,~
A_\psi=\frac{B r^2\cos^2 \theta }{\Lambda},~{\rm with}~\Lambda=1+\frac{d-3}{2(d-2)}B^2 r^2\cos^2 \theta,
\end{eqnarray}
which is a  higher dimensional generalization
of the $d=4$ Melvin magnetic universe \cite{Melvin:1963qx}.
A direct calculation shows that the horizon area and the temperature of the 
new solutions (\ref{h4}), (\ref{h5})
are not affected by the external magnetic field,
coinciding with the corresponding quantities of the $B=0$
seed configurations.

Moreover, by employing the same approach 
as in \cite{Radu:2002pn}, \cite{Ortaggio:2004kr},
it is straightforward
to show that the mass of the new solutions, as defined with respect to the Melvin universe background,
still preserves the expression found in the asymptotically flat case\footnote{The fact that
the thermodynamics of a magnetized static black hole is not affected by the presence 
of the background magnetic field has also  been noticed for other solutions, 
see $e.g.$ \cite{Radu:2002pn}, \cite{Ortaggio:2004kr}, \cite{Yazadjiev:2008ty}.}.

The configurations with generic values of $B$ possess again a conical singularity at $\theta=0$, $r_H<r<R$.
However,  
this conical singularity vanishes for a critical value of the magnetic field,
 \begin{eqnarray}
\label{bc}
B_c= \frac{1}{\cal Q}\frac{4(d-2)}{(d-3)}
\left(1-
(1-\frac{\delta}{2\pi})^{\frac{d-3}{2(d-2)}} 
\right ).
\end{eqnarray}
The dimensionless quantity $b_c=B_c{\mathcal Q}$ is shown in Figure 2 (right) 
as a function of the parameters $a_H$ and $R/r_H$ for $d=5,6,7$ solutions.
One can see that $b_c$ diverges as the Schwarzschild limit is approached.

\section{Conclusions}
In this work we have shown numerical evidence that the vacuum static
black holes with $S^2\times S^{d-4}$
horizon topology discussed in \cite{Kleihaus:2009wh}
admit nonextremal generalizations in Einstein-Maxwell theory.
 These new solutions have a dipolar magnetic field,
which is created by a spherical $S^{d-4}$ distribution of monopoles.
 They also share the basic properties
of the $d=5$ static dipole ring and
possess conical singularities, which, in 
the absence of rotation, prevent the black objects to collapse.
Of course, on general grounds, one expects the $d > 5$ new solutions in this work to possess
rotating generalizations and thus to achieve balance for a critical value of 
the angular momentum.
Unfortunately,  the explicit construction of such solutions proves a very 
 difficult numerical problem, see the discussion in \cite{Kleihaus:2010pr}.

However, as discussed in the second part of this work, 
these  static black objects with a $S^2\times S^{d-4}$ topology of the horizon can be held in equilibrium
by switching on a magnetic field
with an appropriate strength.
To the best of our knowledge, this is the first explicit construction of
$d>5$ static and balanced black objects 
which are regular on and outside an event horizon of non-spherical 
topology\footnote{The topological black holes
in anti-de Sitter spacetime are qualitatively a different class of black objects.}.
However, the magnetic field does not vanish asymptotically, such that the background spacetime
corresponds in this case to a $d-$dimensional Melvin universe.
Therefore the construction of asymptotically flat, static balanced black objects with 
a non-spherical horizon topology remains an open problem.

Our preliminary results indicate that
the solutions in this work can be generalized to include a dilaton.
In this ways, they could be uplifted to higher dimensions and 
interpreted in a string theory context.
 Moreover, we expect that all static configurations with a non-spherical horizon topology discussed in 
 \cite{Kleihaus:2010pr} would admit generalizations with a dipolar magnetic field.
 Although the asymptotically flat static solutions will possess conical singularities, 
 the interaction with an external 
 magnetic field would balance them.

\vspace{0.75cm}
\noindent{\textbf{~~~Acknowledgements.--~}We gratefully acknowledge support by the DFG,
in particular, also within the DFG Research
Training Group 1620 ''Models of Gravity''.

 \begin{small}
 
 \end{small}

 \end{document}